\documentclass[twocolumn,showpacs,preprintnumbers,amsmath,amssymb,superscriptaddress]{revtex4}

\usepackage{graphicx}
\def\para{\,/\!/\,}

\begin{document}

\title{The band structure and Fermi surface of La$_{0.6}$Sr$_{0.4}$MnO$_{3}$ thin films studied by \textit{in-situ} angle-resolved photoemission spectroscopy}

\author{A. Chikamatsu}
\affiliation{Department of Applied Chemistry, The University of Tokyo, 7-3-1 Hongo, Bunkyo-ku, Tokyo 113-8656, Japan}
\author{H. Wadati}
\affiliation{Department of Complexity Science and Engineering and Department of Physics, University of Tokyo, Kashiwa 277-8561, Japan}
\author{H. Kumigashira}
\affiliation{Department of Applied Chemistry, The University of Tokyo, 7-3-1 Hongo, Bunkyo-ku, Tokyo 113-8656, Japan}
\author{M. Oshima}
\affiliation{Department of Applied Chemistry, The University of Tokyo, 7-3-1 Hongo, Bunkyo-ku, Tokyo 113-8656, Japan}
\author{A. Fujimori}
\affiliation{Department of Complexity Science and Engineering and Department of Physics, University of Tokyo, Kashiwa 277-8561, Japan}
\author{N. Hamada}
\affiliation{Department of Physics, Tokyo University of Science, Chiba 278-8510, Japan}
\author{T. Ohnishi}
\affiliation{Institute for Solid State Physics, The University of Tokyo, Kashiwa 277-8581, Japan}
\author{M. Lippmaa}
\affiliation{Institute for Solid State Physics, The University of Tokyo, Kashiwa 277-8581, Japan}
\author{K. Ono}
\affiliation{Institute of Materials Structure Science, High Energy Accelerator Research Organization, Tsukuba 305-0801, Japan}
\author{M. Kawasaki}
\affiliation{Institute for Materials Research, Tohoku University, Sendai 980-8577, Japan}
\author{H. Koinuma}
\affiliation{Materials and Structures Laboratory, Tokyo Institute of Technology,
Yokohama 226-8503, Japan}

\date{\today}

\begin{abstract}
We have performed an \textit{in situ} angle-resolved photoemission spectroscopy (ARPES) on single-crystal surfaces of La$_{0.6}$Sr$_{0.4}$MnO$_{3}$ (LSMO) thin films grown on SrTiO$_{3}$ (001) substrates by laser molecular beam epitaxy, and investigated the electronic structure near the Fermi level ($E_{F}$).  The experimental results were compared with the band-structure calculation based on LDA + $U$.  The band structure of LSMO thin films consists of several highly dispersive O 2$p$ derived bands in the binding energy range of 2.0 - 6.0 eV and Mn 3$d$ derived bands near $E_{F}$.  ARPES spectra around the $\Gamma$ point show a dispersive band near $E_{F}$ indicative of an electron pocket centered at the $\Gamma$ point, although it was not so clearly resolved as an electronlike pocket due to the suppression of spectral weight in the vicinity of $E_{F}$.  Compared with the band-structure calculation, the observed conduction band is assigned to the Mn 3$de_{g}$ majority-spin band responsible for the half-metallic nature of LSMO.  We have found that the estimated size of the Fermi surface is consistent with the prediction of the band-structure calculation, while the band width becomes significantly narrower than the calculation.  Also, the intensity near $E_{F}$ is strongly reduced.  The origin of these discrepancies between the experiment and the calculation is discussed.
\end{abstract}

\pacs{71.18.+y, 71.30.+h, 79.60.-i, 75.47.Lx}

\maketitle

\section{Introduction}
Hole-doped perovskite manganese oxides La$_{1-x}$Sr$_{x}$MnO$_{3}$ (LSMO) have attracted considerable attention because of their interesting magnetic and electrical properties, such as colossal magnetoresistance (CMR) and composition- and temperature-dependent metal-insulator transition \cite{ImadaM:1998, ParkJH:1998}.  In the ferromagnetic metallic phase ($0.17 < x < 0.5$), it has been considered that the system is in the half-metallic state.  The ferromagnetic ground state of LSMO has been discussed in the framework of the double exchange (DE) model \cite{ZenerC:1951, AndersonPW:1955}: holes doped into the Mn 3$d$ band of $e_{g}$ symmetry are considered as mobile charge carriers; they are subject to a strong Hund's coupling to the localized Mn$^{4+}$ (S = 3/2) spins, leading to a fully spin-polarized Mn 3$d$ $e_{g}$-derived conduction band.  However, recent theoretical and experimental studies have revealed that dynamical Jahn-Teller (JT) effect or local lattice distortion also plays an important role in CMR \cite{MillisAJ:1996.2, MillisAJ:1996}.  Precise information about the complicated electronic structure of these materials is therefore necessary for a proper understanding of the unusual physical properties in LSMO.

Angle-resolved photoemission spectroscopy (ARPES) is a unique and powerful experimental technique to determine the band structure of a solid and has long played a central role in the studies of the electronic properties of strongly correlated electron systems \cite{ShenZX:1995}.  However, there have been few ARPES studies on transition metal oxides (TMO) with three-dimensional perovskite structures like LSMO.  This is in sharp contrast to the intensive ARPES studies on high-$T_{C}$ cuprate superconductors with layered structures \cite{JohnsonPD:2001, DamascelliA:2000}.  This is mainly due to the difficulty in obtaining single-crystal surfaces of TMO crystals by standard surface preparation techniques such as cleaving or sputtering and annealing.  The lack of information about their band structures near the Fermi level ($E_{F}$), especially the Fermi surface (FS), has limited the understanding of the physics of the perovskite-type TMO's.

Recent progress in the laser molecular-beam epitaxy (MBE) technique has enabled us to grow TMO films by controlling their growth process on an atomic level \cite{KoinumaH:1997, OhnishiT:2001}.  By using well-defined surfaces of epitaxial films, we have studied the band structures of TMO's and demonstrated that \textit{in situ} ARPES measurements on such TMO films are one of the most ideal methods to investigate the band structure of TMO's with three-dimensional crystal structures \cite{HoribaK:2003, KumigashiraH:2004, ChikamatsuA}.  Recently, the complex band structure near $E_{F}$ of LSMO was reported in an ARPES study of epitaxial LSMO $x$ = 0.34 films, and the coexistence of one-dimensional state along the out-of-plane direction and two- (three-) dimensional band structure was suggested \cite{ShiM:2004}.  In this paper, we report on \textit{in-situ} ARPES results of single-crystal surfaces of LSMO $x$ = 0.4 thin films grown on SrTiO$_{3}$ (001) substrates.  We have clearly observed the Mn 3$de_{g}$ derived conduction band centered at the $\Gamma$ point, which is responsible for the half-metallic behavior of LSMO $x$ = 0.4.  Comparing the present ARPES results with band-structure calculation, we discuss the electronic structure near $E_{F}$ of LSMO.

\section{Experimental}
Experiments were carried out using a photoemission spectroscopy (PES) system combined with a laser MBE chamber, which was installed at beamline BL-1C of the Photon Factory, KEK \cite{HoribaK:2003}.  The LSMO $x$ = 0.4 thin films were grown epitaxially on SrTiO$_{3}$ (STO) substrates by pulsed laser deposition.  Sintered LSMO $x$ = 0.4 pellets were used as ablation targets.  A Nd: YAG laser was used for ablation in its frequency-triple mode ($\lambda$ = 355 nm) at a repetition rate of 1 Hz.  The wet-etched STO (001) substrates \cite{KawasakiM:1994} were annealed at 1050 $^{\circ}$C at an oxygen pressure of 1$\times$10$^{-6}$ Torr before deposition to obtain an atomically flat TiO$_{2}$-terminated surface.  LSMO thin films with a thickness of about 400 {\AA} were deposited on the TiO$_{2}$-terminated STO substrates at a substrate temperature of 1050 $^{\circ}$C and at an oxygen pressure of 1$\times$10$^{-4}$ Torr \cite{KumigashiraH:2003, KumigashiraH:2004.2}.  The intensity of the specular spot in reflection high-energy electron-diffraction (RHEED) patterns was monitored during the deposition to determine the surface morphology and the film growth rate.  Layer-by-layer growth of the LSMO thin films was confirmed by the observation of clear RHEED oscillations.  The thin films were subsequently annealed at 400 $^{\circ}$C for 45 minutes at atmospheric pressure of oxygen to remove oxygen vacancies.  After cooling down below 100 $^{\circ}$C, the films were moved into the photoemission chamber under a vacuum of 10$^{-10}$ Torr.  The ARPES spectra were taken at 25 K using a Gammadata SCIENTA SES-100 electron-energy analyzer in the angle-resolved mode.  The total energy and angular resolution was set at about 150 meV and 0.5$^{\circ}$, respectively.  The Fermi level of the samples was referred to that of a gold foil which was in electrical contact with the sample.  The surface structure and cleanness of the vacuum-transferred LSMO films were checked by low energy electron diffraction (LEED) and core-level photoemission.  The sharp 1$\times$1 spots and some surface reconstruction-derived spots were observed, confirming the well-ordered surface of the films.  The O 1$s$ core level showed a sharp single peak, and no C 1$s$ peak was observed.  These results indicate that no cleaning procedures were needed for the \textit{in situ} ARPES measurement \cite{HoribaK:2003, HoribaK.1}.  The surface morphology of the measured films was analyzed by \textit{ex situ} atomic force microscopy in air and atomically flat step-and-terrace structures were observed. The crystal structure was characterized by four-circle X-ray diffraction (XRD), confirming coherent growth of the films on the substrates.  The Curie temperature of measured films was determined to be 354 K using a superconducting quantum interference device (SQUID) magnetometer, nearly the same as the reported values \cite{HoribaK.1, IzumiM:1998}.   

\section{Results and discussion}
Figure~1 shows ARPES spectra of an LSMO thin film measured at 25 K with photon energy of 88 eV along the "$\Gamma$-X direction" in the tetragonal Brillouin-zone (BZ) of an epitaxial LSMO thin film.  The polar angle ($\theta$) was measured from the (001) surface normal.  Due to the short escape depth of photoelectrons, however, there is substantial broadening of momentum of photoelectrons perpendicular to the surface ($k_{z}$), while the momentum parallel to the surface is conserved.  Under such conditions, the ARPES spectra are integrated along the $k_{z}$ direction over $\sim$20 \% of the Brillouin zone and therefore reflect band dispersions along the high-symmetry line (a $\Gamma$-X line in the present case) as a result of relatively large density of states on the high-symmetry line even though the momentum determined in a free-electron final state model deviates from high-symmetry lines.  The ARPES spectra exhibit considerable and systematic changes as a function of the polar angle; highly dispersive features with a dispersional band width of about 2 eV exist in the binding energy range of 2.0 - 6.0 eV, while almost dispersionless features are observed at about 1.5 eV.  In the vicinity of the Fermi level, one notices a small but distinct dispersive feature around the binding energy of 0.5 eV in the region of $\theta$ = $-3.6^{\circ}$ - $3.6^{\circ}$ and $16.4^{\circ}$ - $23.6^{\circ}$ (i.e. around the $\Gamma$ point), suggesting the existence of an electron pocket at the center of BZ.
\begin{figure}
\begin{center}
\includegraphics[width=8.5cm]{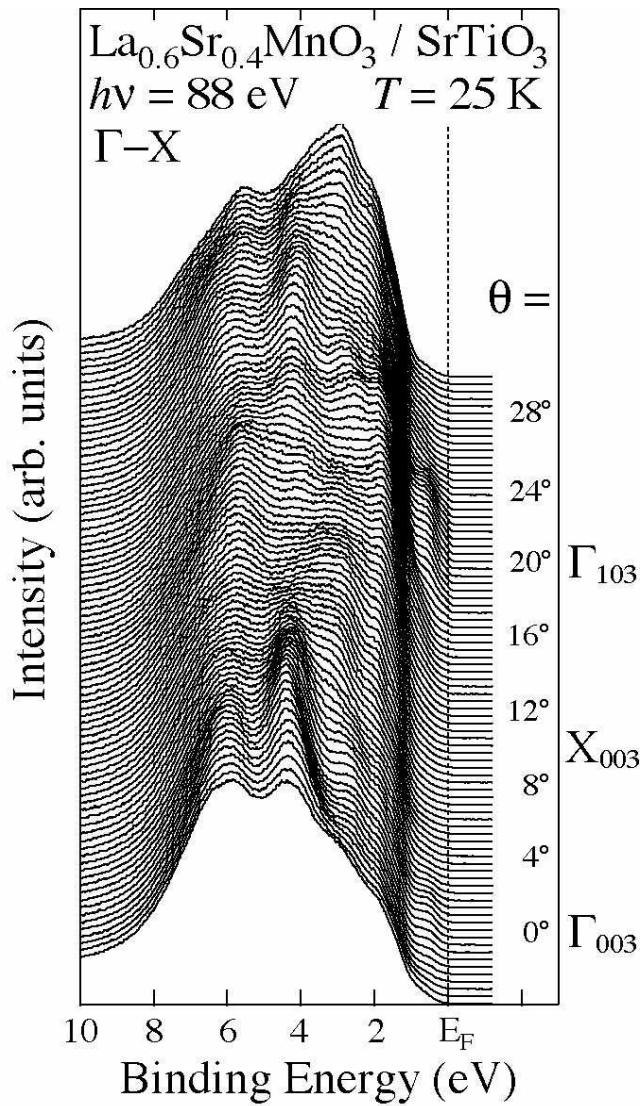}
\caption{\textit{In-situ} angle-resolved photoemission spectra along $\Gamma$-X direction of a La$_{0.6}$Sr$_{0.4}$MnO$_{3}$ thin film grown epitaxially on a SrTiO$_{3}$ substrate by laser MBE.  Polar angle ($\theta$) referred to the surface normal is indicated.}
\label{fig1}
\end{center}
\end{figure}

In order to see more clearly the dispersive feature of the bands in the ARPES spectra, we have mapped out the experimental gband structureh in Fig.~2(a).  Here, the band structure has been obtained by taking the second derivative of the ARPES spectra after smoothing and plotting the intensity in the gray scale \cite{HimpselFJ:1983}.  Dark parts correspond to the energy bands.  As expected from ARPES spectra shown in Fig.~1, several highly dispersive bands at higher binding energies and almost dispersionless structures at about 1.5 eV are clearly observed.  The experimental band structure is almost symmetric with respect to the high-symmetry points in the BZ of the LSMO thin film.  In Fig.~2(a), we find that each band at the equivalent $k$ points ($\Gamma_{003}$ and $\Gamma_{103}$ points) has almost the same energy levels.  This means that the perpendicular component of the photoelectron momentum is broadened.  It should be noted that we also measured ARPES spectra with varying the photon energy, and confirmed that the energy positions are almost identical when the deviation of $k$-perpendicular from the $\Gamma$-X high-symmetry line is within $\pm$20 \% of the reciprocal lattice length along $k$-perpendicular direction.

	Next, we compare the experimental band structure obtained by ARPES with the band-structure calculation for LSMO based on the local density approximation (LDA) including on-site Coulomb interactions $U$ (Fig.~2(b)).  The calculation is performed with $U$ = 2 eV using the virtual crystal approximation for the La/Sr site to account for the doping level of 0.4 holes per Mn site.  The tetragonal crystal structure of the LSMO films due to tensile strain given by STO substrates is also taken into account (The a-axis length $d_{\para}$ of 0.391 nm and the c-axis length $d_{\bot}$ of 0.383 nm, which are determined by the XRD measurements, are used.).  The space of the electronic states is separated into the subspace of the localized orbitals, for which the Coulomb interaction between electrons is explicitly taken into account, and the subspace of the delocalized states, for which the orbital independent Kohn-Sham one electron potential is considered to be a good approximation \cite{AnisimovVI:1993, SolovyevIV:1994, SawadaH:1997}.  The solid and dashed lines correspond to the majority and minority bands, respectively.  As found in Fig.~2, the overall feature of the experimental band structure shows qualitatively good agreement with the LDA + $U$ band-structure calculation.  According to the band-structure calculation, several highly dispersive bands at higher binding energies originate mainly from the O 2$p$ dominant states, while the dispersionless band located at about 1.5 eV is assigned to the Mn 3$d$ states.  We also find a small portion of a dispersive band near $E_{F}$ at the $\Gamma$ point, which may be ascribed to the electron pocket at the $\Gamma$ point predicted by the band-structure calculation.
\begin{figure*}
\begin{center}
\includegraphics[width=15cm]{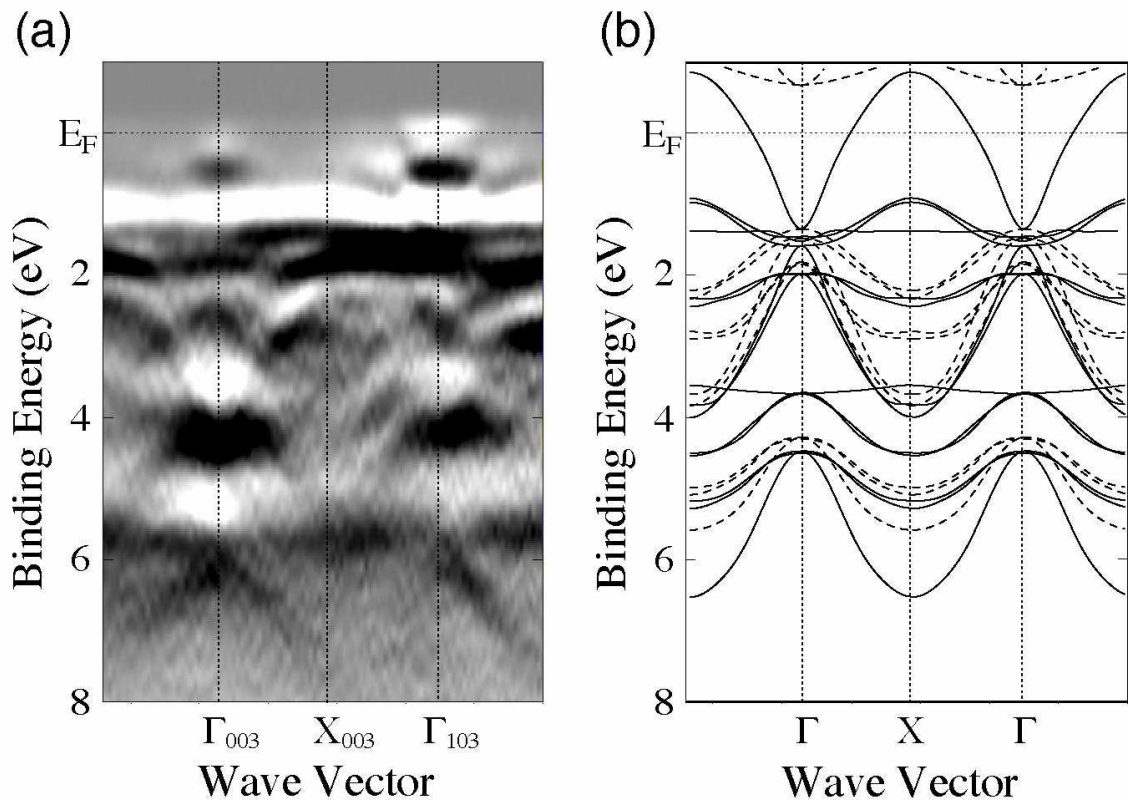}
\caption{(a) Experimental band structure in the $\Gamma$-X direction of the La$_{0.6}$Sr$_{0.4}$MnO$_{3}$ thin film obtained from the ARPES measurements.  Dark parts correspond to the energy bands.  (b) The band-structure calculation for LSMO using the local density approximation including on-site Coulomb interactions (LDA + $U$, $U$ = 2.0 eV).  Solid and broken lines correspond to majority and minority bands, respectively.}
\label{fig2}
\end{center}
\end{figure*}

   @ Since the electron FS is fully spin polarized and is responsible for the half metallic state according to the calculation, it is essential to investigate the behavior of the dispersive band near $E_{F}$ and its $E_{F}$ crossing in more detail.  We show the ARPES spectra around the $\Gamma_{103}$ point in the near-$E_{F}$ region as shown in Fig.~3 on an expanded scale.  The dispersion is more clearly seen around the $\Gamma_{103}$ point than in the equivalent $\Gamma_{003}$ point owing mainly to matrix-element effects.  In Fig.~3, one finds at least two dispersive bands near $E_{F}$ (denoted by A and B).  The dispersion of band B is more clearly seen in the band mapping of the second-derivative ARPES spectra as shown in Fig.~2(a) or Fig.~5(a).  The dispersive band with relatively broad ARPES peaks in the energy range of 0.5 - 1.2 eV has been reported in the recent ARPES measurements in both normal emission geometry with varying the photon energy and off normal geometry along different lines in the BZ \cite{ShiM:2004}.  The observation of the broad features in different $k$ points of the BZ indicates that band B originates from the occupied band below the binding energy of 0.5 eV.  On the other hand, the dispersive band A around $\theta$ = 20$^{\circ}$ exhibits symmetric energy dispersion with respect to $\theta$ = 20$^{\circ}$ as shown by filled or opened triangles.  It has a bottom at 0.5-eV binding energy at $\theta$ = 19.6 - 20.4$^{\circ}$ and gradually approaches $E_{F}$ with increasing or decreasing polar angle.  This suggests that band A crosses $E_{F}$ and enters the unoccupied states around $\theta$ = 16.4$^{\circ}$ (and 23.6$^{\circ}$) owing to the strong suppression in its spectral intensity as it approaches $E_{F}$.  The sudden appearance of spectral weight at $E_{F}$ around 16-18$^{\circ}$ with increasing $\theta$ (around 22-24$^{\circ}$ with increasing $\theta$) can be attributed to the $E_{F}$ crossing.  The Fermi edge emerges in the ARPES spectra at $\theta$ = 16.4$^{\circ}$ with increasing $\theta$ and persists up to $\theta$ = 23.6$^{\circ}$.  The existence of a Fermi edge responsible for the metallic nature of LSMO is clearly seen by comparing the ARPES spectra with the gold Fermi edge as shown in Fig.~4(a).  In contrast, the ARPES spectra outside the electron pocket exhibit a concave line shape around $E_{F}$ and no Fermi cutoff, suggesting the crossing of band A to $E_{F}$.  The appearance of the Fermi edge inside the ``electron pocket'' is derived from the finite momentum broadening along $k$-perpendicular due to final states effect as described above \cite{GrandkeT:1978}.  In order to estimate the Fermi momentum ($k_{F}$) from the ARPES experiment in more detail, we plot the angular distribution function ($n(\theta)$) and the absolute value of its differential curve ($|dn(\theta)/d\theta |$) in Fig.~4(b), where $n(\theta)$ is defined as the ARPES spectral intensity near $E_{F}$ integrated from $-150$ meV to $150$ meV with respect to $E_{F}$ .  We have estimated $k_{F}$ to be $-0.25$ {\AA$^{-1}$} and $0.27$ {\AA$^{-1}$} ($\theta$ = 16.8$^{\circ}$ and 23.6$^{\circ}$) measured from $\Gamma_{103}$ point by taking the maximum point of $|dn(\theta)/d\theta |$ \cite{StraubT:1997}.  The values are in good agreement with $k_{F}$'s determined by extrapolating the energy position of the dispersive feature in the ARPES spectra.  From these results, it is inferred that the band calculation serves as a good approximation in describing the Fermi surface of LSMO.  
\begin{figure}
\begin{center}
\includegraphics[width=8.5cm]{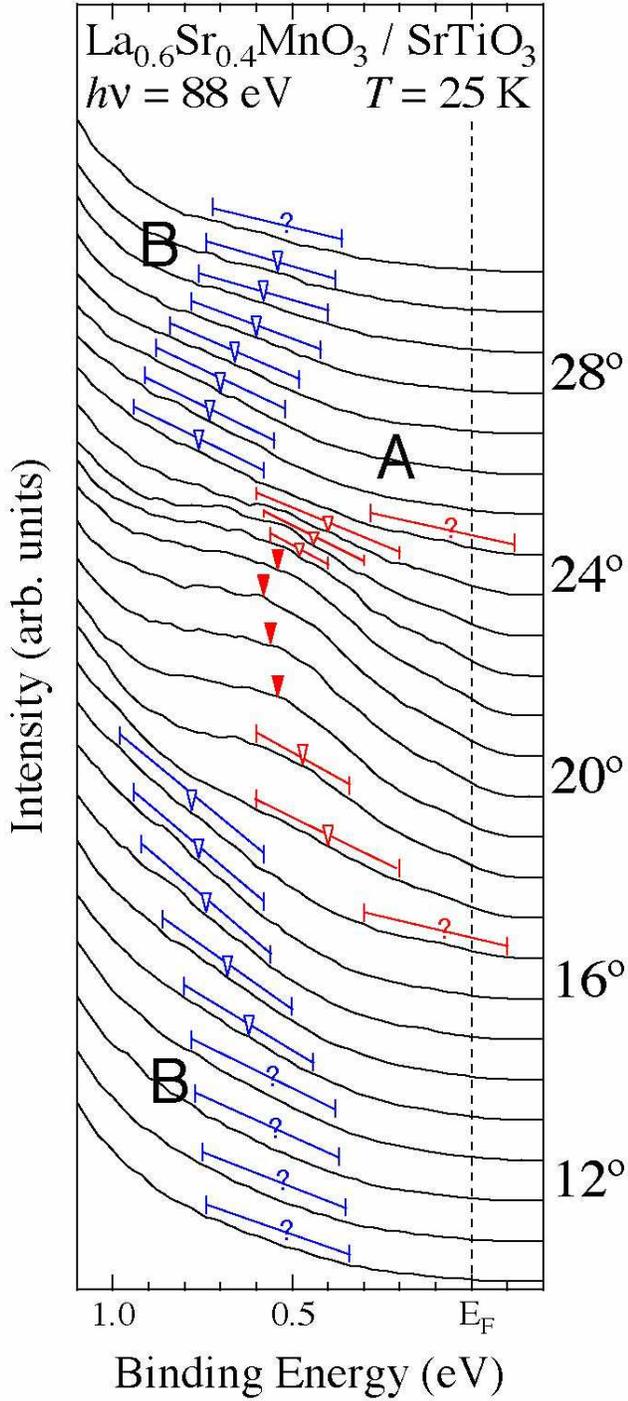}
\caption{The ARPES spectra near $E_{F}$ of a La$_{0.6}$Sr$_{0.4}$MnO$_{3}$ thin film in an enlarged binding energy scale measured around the $\Gamma_{103}$ point.  ARPES spectra show two bands: dispersive band A and relatively weak dispersive band B.  Solid and open triangles and question marks represent the energy positions of prominent, medium, and weak structures, respectively.  The energy positions are determined by the second derivative spectra.}
\label{fig3}
\end{center}
\end{figure}
\begin{figure}
\begin{center}
\includegraphics[width=8.5cm]{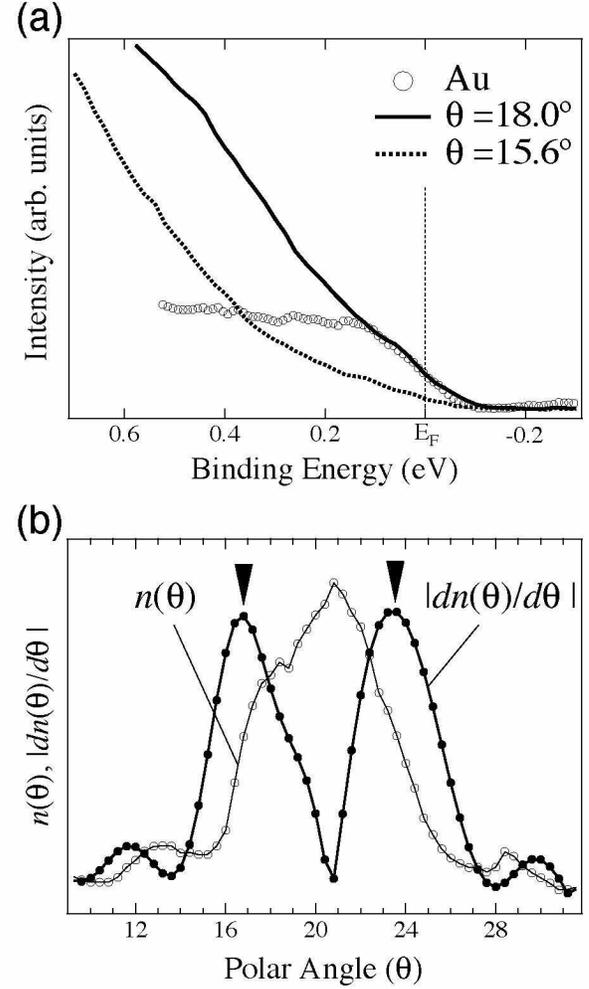}
\caption{(a) ARPES spectra of $\theta$ = 18$^{\circ}$ (inside the ``electron pocket'') and $\theta$ = 15.6$^{\circ}$ (outside the ``electron pocket''), together with photoemission spectrum of gold taken with almost the same experimental conditions. (b) An angular distribution curve ($n(\theta)$) obtained by plotting the ARPES intensity integrated from $-150$ meV to $+150$ meV with respect to $E_{F}$ as a function of $\theta$, together with absolute value of its differential curve ($|dn(\theta)/d\theta |$).  Arrows represent the maximum of $|dn(\theta)/d\theta |$ indicative of the Fermi momenta.}
\label{fig4}
\end{center}
\end{figure}

Figure~5 shows comparison of the experimental band structure near $E_{F}$ obtained from the present ARPES spectra with the LDA+$U$ band-structure calculation.  The experimental band structure shows qualitatively good agreement with the band-structure calculation.  In the calculation, the band forming the electron FS is ascribed to the Mn 3$d(3x^{2}-r^{2})$ states, while the weakly dispersive bands below the binding energy of 0.9 eV have substantial Mn 3$dt_{2g}$ character.  Besides the dispersive bands, there is a flat band at the binding energy of 1.5 eV, which is assigned to the Mn-3$d$ band originated in $y^{2}-z^{2}$ orbital (see also Fig.~2).  While the 3$d(y^{2}-z^{2})$ derived band shows excellent agreement between the experiment and the calculation, the experimental bands near $E_{F}$ are substantially different from the band-structure calculation; the energy position of band B at the X point shifts to higher binding energy by about 0.5 eV, and the bottom of the conduction band (band A), which forms an electron FS centered at the $\Gamma$ point in the calculation, also shifts to the higher binding energy by 0.8 eV. 
\begin{figure}
\begin{center}
\includegraphics[width=8.5cm]{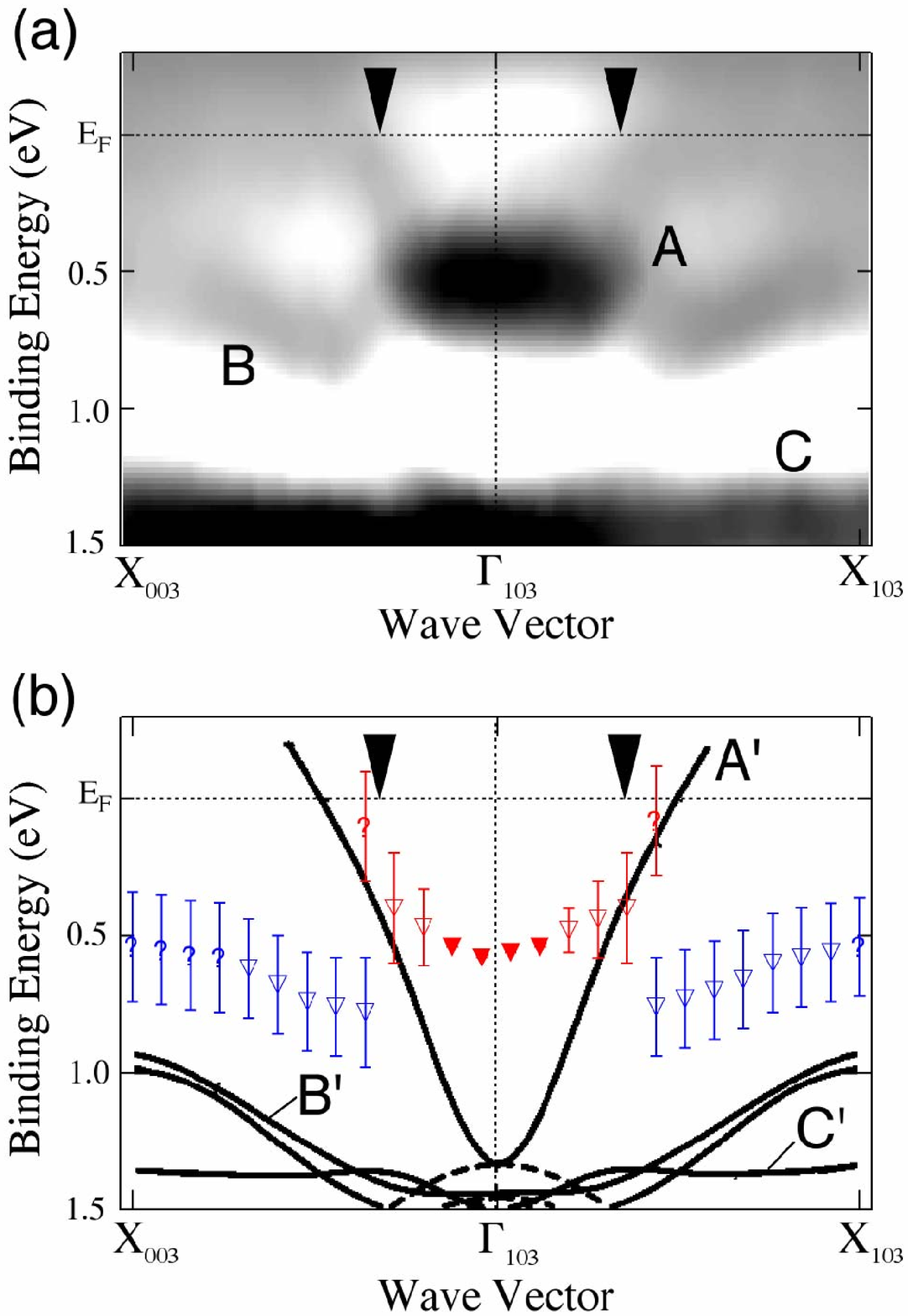}
\caption{Comparison of the band structure near $E_{F}$ of LSMO thin films around the $\Gamma_{103}$ point between (a) experimental band structure near $E_{F}$ determined by the present ARPES and (b) the band-structure calculation.  The peak positions of the band A and B in the ARPES spectra (Fig.~3) are superimposed in the calculation with red and blue marks, respectively.  Solid and open triangles and question marks represent prominent, medium, and weak structure in the ARPES spectra shown in Fig.~3, respectively.  The Fermi momenta determined by ADC (Fig.~4(b)) are also shown by arrows.  In the calculation, the bands labeled by A' and C' are assigned to the Mn 3$de_{g}$ states originated in 3$x^{2}-r^{2}$ and $y^{2}-z^{2}$ orbitals, respectively, while the band B' is ascribed to be Mn 3$dt_{2g}$ states.  Here, $z$-axis direction is defined as the direction perpendicular to the film surface, while ARPES measurement is performed along $x$-axis direction.}
\label{fig5}
\end{center}
\end{figure}
 
The possible explanation for the upward shift of conduction-band minima is the renormalization effect due to Coulomb interaction.  While the experimental $k_{F}$ and the calculated $k_{F}$ agree with each other in LSMO, the considerable narrowing of the conduction band suggests the importance of renormalization effect due to the strong electron correlation, which is not included in the band calculation.  The vanishing or very small spectral weight at $E_{F}$ and the anomalously broad ARPES spectral features shows close similarity to the previous ARPES spectra of layered manganites \cite{DessauDS:1998, SaitohT:2000, ChuangYD:2001}.  The significant suppression of ARPES spectral weight in the near-$E_{F}$ region is also reminiscent of the pseudogap formaion at $E_{F}$ in angle-integrated PES \cite{HoribaK.1, SaitohT:1995} as well as the strong suppression of Drude peak in the optical measurements \cite{ArimaT:1993, OkimotoY:1995, OkimotoY:1997}, while weak but distinct Fermi-edge cutoff is clearly observed in three-dimensional perovskite manganites \cite{HoribaK.1, SarmaDD:1996}.  

The presence of the pseudogap at $E_{F}$ seems to be a phenomenon in PES spectra of two- and three-dimensional manganites.  It is reported in the ARPES study of layered manganites that the pseudogap originates from charge density wave (CDW) in cooperation with the JT distortion, since the nesting vector estimated from the ARPES measurements is in line with an additional superlattice reflection observed in x-ray scattering experiments on the layered materials \cite{ChuangYD:2001, KubotaM:2000}.  However, such a CDW instability is hardly expected in the three-dimensional materials.  In fact, the calculated FS has no parallel FS portion responsible for CDW instability.  This argument is further supported by the absence of any superlattice spots in manganites with a three-dimensional perovskite structure \cite{KubotaM:2000}.  These results suggest that the coupling of electrons to a local mode such as JT distortion may play a dominant role in the pseudogap behavior of the ARPES spectra of LSMO.  In order to investigate the unusual behaviors of the ARPES spectra of manganites, systematic ARPES studies as a function of hole concentration and/or temperature are needed.

\section{Conclusion}
We have performed an \textit{in situ} angle-resolved photoemission spectroscopy (ARPES) on single-crystal surfaces of La$_{0.6}$Sr$_{0.4}$MnO$_{3}$ (LSMO) thin films to investigate the complicated electronic structure near the Fermi level ($E_{F}$).  The valence-band structure consisting of the O 2$p$ and Mn 3$d$ states shows a qualitatively good agreement between the ARPES experiment and band-structure calculation based on LDA+$U$.  ARPES spectra near $E_{F}$ around the $\Gamma$ point show a dispersive band near $E_{F}$ indicative of an electron pocket centered at the $\Gamma$ point, though it is not so clearly resolved as an electron pocket due to the strong suppression of spectral weight in the vicinity of $E_{F}$.  Compared with the LDA+$U$ band-structure calculation, the dispersive conduction band is assigned to the Mn 3$de_{g}$ majority band responsible for the half-metallic nature of LSMO.  We have found that the estimated size of the electron FS closely agrees with the band-structure calculation, while the width of conduction band is significantly renormalized by the many-body interactions, reflecting the unusual physical properties of the manganites.

\section{Acknowledgment}
We are very grateful to A. Chainani, K. Horiba, M. Kubota, and J. Matsuno for useful discussion.  This work was done under Project No. 02S2-002 at the Institute of Materials Structure Science at KEK.  This work was supported by a Grant-in-Aid for Scientific Research (A16204024) from the Japan Society for the Promotion of Science.

\bibliography{LSMO04_ARPES}

\begin{thebibliography}{10}

\bibitem{ImadaM:1998}
M. Imada, A. Fujimori, and Y. Tokura, Rev. Mod. Phys. {\bf 70},  1039  (1998),
  and references there in.

\bibitem{ParkJH:1998}
J.-H. Park, E. Vescovo, H.-J. Kim, C. Kwon, R. Ramesh, and T. Venkatesan,
  Nature {\bf 392},  794  (1998).

\bibitem{ZenerC:1951}
C. Zener, Phys. Rev. {\bf 82},  403  (1951).

\bibitem{AndersonPW:1955}
P.~W. Anderson and H. Hasegawa, Phys. Rev. {\bf 100},  675  (1955).

\bibitem{MillisAJ:1996.2}
A.~J. Millis, B.~I. Shraiman, and R. Mueller, Phys. Rev. Lett. {\bf 77},  175
  (1996).

\bibitem{MillisAJ:1996}
A.~J. Millis, R. Mueller, and B.~I. Shraiman, Phys. Rev. B {\bf 54},  5405
  (1996).

\bibitem{ShenZX:1995}
Z.-X. Shen and D.~S. Dessau, Phys. Rep. {\bf 253},  1  (1995).

\bibitem{JohnsonPD:2001}
P.~D. Johnson, T. Valla, A.~V. Fedorov, Z. Yusof, B.~O. Wells, Q. Li, A.~R.
  Moodenbaugh, G.~D. Gu, N. Koshizuka, C. Kendziora, S. Jian, and D.~G. Hinks,
  Phys. Rev. Lett. {\bf 87},  177007  (2001).

\bibitem{DamascelliA:2000}
A. Damascelli, D.~H. Lu, K.~M. Shen, N.~P. Armitage, F. Ronning, D.~L. Feng, C.
  Kim, Z.-X. Shen, T. Kimura, Y. Tokura, Z.~Q. Mao, and Y. Maeno, Phys. Rev.
  Lett. {\bf 85},  5194  (2000).

\bibitem{KoinumaH:1997}
H. Koinuma, N. Kanda, J. Nishio, A. Ohtomo, H. Kubota, M. Kawasaki, and M.
  Yoshimoto, Appl. Surf. Sci. {\bf 109/110},  514  (1997).

\bibitem{OhnishiT:2001}
T. Ohnishi, D. Komiyama, T. Koida, S. Ohashi, C. Stauter, H. Koinuma, A.
  Ohtomo, M. Lippmaa, N. Nakagawa, M. Kawasaki, T. Kikuchi, and K. Omote, Appl.
  Phys. Lett. {\bf 79},  536  (2001).

\bibitem{HoribaK:2003}
K. Horiba, H. Ohguchi, H. Kumigashira, M. Oshima, K. Ono, N. Nakagawa, M.
  Lippmaa, M. Kawasaki, and H. Koinuma, Rev. Sci. Instr. {\bf 74},  3406
  (2003).

\bibitem{KumigashiraH:2004}
H. Kumigashira, K. Horiba, H. Ohguchi, M. Oshima, N. Nakagawa, M. Lippmaa, K.
  Ono, M. Kawasaki, and H. Koinuma, J. Magn. Magn. Mater. {\bf 272-276},  434
  (2004).

\bibitem{ChikamatsuA}
A. Chikamatsu, H. Wadati, M. Takizawa, R. Hashimoto, H. Kumigashira, M. Oshima,
  A. Fujimori, N. Hamada, T. Ohnishi, M. Lippmaa, K. Ono, M. Kawasaki, and H.
  Koinuma, J. Electron Spectrosc. Relat. Phenom.  , in press.

\bibitem{ShiM:2004}
M. Shi, M.~C. Falub, P.~R. Willmott, J. Krempasky, R. Herger, K. Hricovini, and
  L. Patthey, Phys. Rev. B {\bf 70},  140407(R)  (2004).

\bibitem{KawasakiM:1994}
M. Kawasaki, K. Takahashi, T. Maeda, R. Tsuchiya, M. Shinohara, O. Ishihara, T.
  Yonezawa, M. Yoshimoto, and H. Koinuma, Science {\bf 266},  1540  (1994).

\bibitem{KumigashiraH:2003}
H. Kumigashira, K. Horiba, H. Ohguchi, K. Ono, M. Oshima, N. Nakagawa, M.
  Lippmaa, M. Kawasaki, and H. Koinuma, Appl. Phys. Lett. {\bf 82},  3430
  (2003).

\bibitem{KumigashiraH:2004.2}
H. Kumigashira, D. Kobayashi, R. Hashimoto, A. Chikamatsu, M. Oshima, N.
  Nakagawa, T. Ohnishi, M. Lippmaa, H. Wadati, K. Ono, M. Kawasaki, and H.
  Koinuma, Appl. Phys. Lett. {\bf 84},  5353  (2004).

\bibitem{HoribaK.1}
K. Horiba, A. Chikamatsu, H. Kumigashira, M. Oshima, N. Nakagawa, M. Lippmaa,
  K. Ono, M. Kawasaki, and H. Koinuma, cond-mat/0406315  .

\bibitem{IzumiM:1998}
M. Izumi, Y. Konishi, T. Nishihara, S. Hayashi, M. Shinohara, M. Kawasaki, and
  Y. Tokura, Appl. Phys. Lett. {\bf 73},  2497  (1998).

\bibitem{HimpselFJ:1983}
F.~J. Himpsel, Adv. Phys. {\bf 32},  1  (1983).

\bibitem{AnisimovVI:1993}
V.~I. Anisimov, I.~V. Solovyev, M.~A. Korotin, M.~T. Czyzyk, and G.~A.
  Sawatzky, Phys. Rev. B {\bf 48},  16929  (1993).

\bibitem{SolovyevIV:1994}
I.~V. Solovyev, P.~H. Dederichs, and V.~I. Anisimov, Phys. Rev. B {\bf 50},
  16861  (1994).

\bibitem{SawadaH:1997}
H. Sawada, Y. Morikawa, K. Terakura, and N. Hamada, Phys. Rev. B {\bf 56},
  12154  (1997).

\bibitem{GrandkeT:1978}
T. Grandke, L. Ley, and M. Cardona, Phys. Rev. B {\bf 18},  3847  (1978).

\bibitem{StraubT:1997}
T. Straub, R. Claessen, P. Steiner, S. H{\"{u}}fner, V. Eyert, K. Friemelt, and
  E. Bucher, Phys. Rev. B {\bf 55},  13473  (1997).

\bibitem{DessauDS:1998}
D.~S. Dessau, T. Saitoh, C.-H. Park, Z.-X. Shen, P. Villella, N. Hamada, Y.
  Moritomo, and Y. Tokura, Phys. Rev. Lett. {\bf 81},  192  (1998).

\bibitem{SaitohT:2000}
T. Saitoh, D.~S. Dessau, Y. Moritomo, T. Kimura, Y. Tokura, and N. Hamada,
  Phys. Rev. B {\bf 62},  1039  (2000).

\bibitem{ChuangYD:2001}
Y.-D. Chuang, A.~D. Gromko, D.~S. Dessau, T. Kimura, and Y. Tokura, Science
  {\bf 292},  1509  (2001).

\bibitem{SaitohT:1995}
T. Saitoh, A.~E. Bocquet, T. Mizokawa, H. Namatame, A. Fujimori, M. Abbate, Y.
  Takeda, and M. Takano, Phys. Rev. B {\bf 51},  13942  (1995).

\bibitem{ArimaT:1993}
T. Arima, Y. Tokura, and J.~B. Torrance, Phys. Rev. B {\bf 48},  17006  (1993).

\bibitem{OkimotoY:1995}
Y. Okimoto, T. Katsufuji, T. Ishikawa, A. Urushibara, T. Arima, and Y. Tokura,
  Phys. Rev. Lett. {\bf 75},  109  (1995).

\bibitem{OkimotoY:1997}
Y. Okimoto, T. Katsufuji, T. Ishikawa, T. Arima, and Y. Tokura, Phys. Rev. B
  {\bf 55},  4206  (1997).

\bibitem{SarmaDD:1996}
D.~D. Sarma, N. Shanthi, S.~R. Krishnakumar, T. Saitoh, T. Mizokawa, A.
  Sekiyama, K. Kobayashi, A. Fujimori, E. Weschke, R. Meier, G. Kaindl, Y.
  Takeda, and M. Takano, Phys. Rev. B {\bf 53},  6873  (1996).

\bibitem{KubotaM:2000}
M. Kubota, Y. Oohara, H. Yoshizawa, H. Fujioka, K. Shimizu, K. Hirota, Y.
  Moritomo, and Y. Endoh, J. Phys. Soc. Jpn. {\bf 69},  1986  (2000).

\end{thebibliography}

\end{document}